\documentclass[12pt]{article}
\usepackage{amsmath}

\csname @addtoreset\endcsname{equation}{section}
\newcommand{\eq}{\begin{equation}}
\newcommand{\feq}{\end{equation}}
\newcommand{\eqn}{\begin{eqnarray}}
\newcommand{\feqn}{\end{eqnarray}}
\newcommand{\arr}{\begin{eqnarray*}}
\newcommand{\farr}{\end{eqnarray*}}

\font\mybb=msbm10 at 12pt
\def\bb#1{\hbox{\mybb#1}}

\def\bR {\bb{R}}

\begin{document}

\begin{titlepage}
\begin{flushright}
IFUM 667/FT\\
CAMS/00-10\\
hep-th/0011016
\end{flushright}
\vspace{.3cm}
\begin{center}
\renewcommand{\thefootnote}{\fnsymbol{footnote}}
{\Large \bf General (Anti-)de~Sitter Black Holes in Five Dimensions}
\vfill
{\large \bf {D.~Klemm$^1$\footnote{email: dietmar.klemm@mi.infn.it} and
W.~A.~Sabra$^2$\footnote{email: ws00@aub.edu.lb}}}\\
\renewcommand{\thefootnote}{\arabic{footnote}}
\setcounter{footnote}{0}
\vfill
{\small
$^1$ Dipartimento di Fisica dell'Universit\`a di Milano
and\\ INFN, Sezione di Milano,
Via Celoria 16,
20133 Milano, Italy.\\
\vspace*{0.4cm}
$^2$ Center for Advanced Mathematical Sciences (CAMS)
and\\
Physics Department, American University of Beirut, Lebanon.\\}
\end{center}
\vfill
\begin{center}
{\bf Abstract}
\end{center}
We find a general class of rotating charged black hole solutions to $N=2$,
$D=5$ gauged supergravity coupled to vector supermultiplets.
The supersymmetry properties of these solutions are studied, and
their mass and angular momenta are obtained. We also compute the
stress tensor of the dual $D=4$, ${\cal N}=4$ super Yang-Mills theory
in the limit of strong `t Hooft coupling. It is shown that closed
timelike curves occur outside the horizon, indicating 
loss of unitarity in the dual CFT.
For imaginary coupling constant of the gravitini to the gauge fields,
one can obtain multi-centered rotating charged de~Sitter black holes.
Some physical properties of these solutions are also discussed.

\end{titlepage}

\section{Introduction}

\label{intro}

Since the advent of the duality between string theory (or, for energies much
smaller than the string scale, supergravity) on anti-de~Sitter (AdS) spaces
and supersymmetric conformal field theories residing on the conformal
boundary of AdS \cite{adscft}, there has been a renewed interest in the
study of AdS black holes. The reason for this is mainly based on the
possibility to study the nonperturbative structure of these field theories
by means of classical supergravity solutions. One hope is that this
correspondence will eventually lead to a deeper insight into confinement in
QCD, or at least into the strong coupling regime of gauge theories with
fewer supersymmetries. A standard example for the exploration of the
nonperturbative regime of conformal field theories using supergravity
solutions is the Hawking-Page phase transition \cite{hawkpage}, which can be
interpreted as a thermal phase transition from a confining to a deconfining
phase in the dual $D=4$, $\mathcal{N}=4$ super Yang-Mills theory \cite
{witten}. Up to now, the study of AdS black holes in the context of the
AdS/CFT correspondence has been extended in various directions \cite
{chamblin,emparan99,caldklemm,awad,awad2,hht,mirjam}, and several interesting
observations were made, like the striking resemblance of the phase structure
of Reissner-Nordstr\"{o}m-AdS black holes to that of a van der Waals-Maxwell
liquid-gas system \cite{chamblin}, the possible appearance of so-called
''precursor'' states \cite{susskind} in the CFT duals of hyperbolic black
holes \cite{emparan99}, or the disappearance of the Hawking-Page transition
for finite `t Hooft coupling \cite{caldklemm}. Besides, by considering
Kerr-AdS black holes, the authors of \cite{awad2} found an explicit example
of a field theory which is scale invariant but not conformally invariant.
Finally, it is worth mentioning that recently there has been an interest
into black strings in AdS space \cite{chamsabra99,klemmsabra00},
because these describe a
flow of four-dimensional $\mathcal{N}=4$ SYM theory on $\bR^{2}\times \Sigma
_{g}$ (where $\Sigma _{g}$ denotes a genus $g$ Riemann surface) to a
two-dimensional CFT in the infrared \cite{klemmsabra00,nunez}.\newline
These results motivate a further research in this field, in particular it is
of interest to find new solutions to various gauged supergravity theories.
Black holes which preserve some of the original supersymmetries could
correspond to an expansion around nonzero vacuum expectation values of
certain CFT operators. In this paper we will be mainly concerned in studying
charged rotating solutions to the theory of $N=2$, $D=5$ gauged supergravity
coupled to an arbitrary number of abelian vector multiplets. Such solutions
were hitherto unknown.\newline
This paper is organized as follows: In section \ref{sugra} we briefly review
$N=2$, $D=5$ gauged supergravity theories coupled to vector multiplets. In
section \ref{bpssol} we derive general solutions for these theories which
admit Killing spinors. In section \ref{rotsymm} black holes with rotational
symmetry in two orthogonal planes are explicitly constructed, and in section 
\ref{physdisc} we discuss the physics of the obtained solutions,
i.~e.~we compute their mass and angular momenta, and show that there exist
closed timelike curves in the region outside the horizon. The possible
relationship of this causality violation with loss of unitarity in
the dual conformal field theory is also discussed.
Furthermore, the stress tensor of this CFT is
determined. Finally, section \ref{desitter} is devoted to the de~Sitter
counterparts obtained by analytically continuing to imaginary coupling
constant of the gravitini. In this case, a Smarr formula for the black holes
is derived. We conclude with some final remarks.

\section{$N=2$, $D=5$ Gauged Supergravity}

\label{sugra}

The theory we shall deal with is $N=2$, $D=5$ gauged supergravity coupled to
an arbitrary number $n$ of abelian vector supermultiplets. The bosonic part
of the Lagrangian is given by 
\begin{equation}
e^{-1}\mathcal{L}=\frac{1}{2}R+g^{2}V-\frac{1}{4}G_{IJ}F_{\mu \nu
}^{I}F^{J\mu\nu}-\frac{1}{2}\mathcal{G}_{ij}\partial _{\mu }\phi
^{i}\partial ^{\mu }\phi ^{j}+\frac{e^{-1}}{48}\epsilon ^{\mu \nu \rho
\sigma \lambda }C_{IJK}F_{\mu \nu }^{I}F_{\rho \sigma }^{J}A_{\lambda }^{K},
\label{action}
\end{equation}
where $\mu ,\nu $ are spacetime indices, $R$ is the scalar curvature, $%
F_{\mu \nu }^{I}$ denote the abelian field-strength tensors, and $e$ is the
determinant of the F\"{u}nfbein $e_{\mu }^{a}$. The scalar potential $V$ is
given by 
\begin{equation}
V(X)=V_{I}V_{J}\left( 6X^{I}X^{J}-\frac{9}{2}\mathcal{G}^{ij}\partial
_{i}X^{I}\partial _{j}X^{J}\right) .  \label{pot}
\end{equation}
The $X^{I}$ are functions of the real scalar fields subject to the condition 
$\mathcal{V}=\frac{1}{6}C_{IJK}X^{I}X^{J}X^{K}=1$. The gauge and the scalar
couplings are determined by the homogeneous cubic polynomial
$\mathcal{V}$ which defines a ''very special geometry''. They are given by 
\begin{eqnarray}
G_{IJ} &=&-\frac{1}{2}\partial _{I}\partial _{J}\log \mathcal{V}\Big|_{%
\mathcal{V}=1},  \notag \\
\mathcal{G}_{ij} &=&\partial _{i}X^{I}\partial _{j}X^{J}G_{IJ}\Big|_{%
\mathcal{V}=1},
\end{eqnarray}
where $\partial _{i}$ and $\partial _{I}$ refer, respectively, to partial
derivatives with respect to the scalar fields $\phi ^{i}$ and $%
X^{I}=X^{I}(\phi ^{i})$.

Furthermore, the constants $V_I$ that arise in (\ref{pot}) specify the
appropriate linear combination of the vectors that comprise the $N=2$
graviphoton, $\mathcal{A}_{\mu} = V_I A_{\mu}^I$.

For Calabi-Yau compactifications of M-theory, $\mathcal{V}$ denotes the
intersection form, and $X^{I}$ and $X_{I}\equiv \frac{1}{6}C_{IJK}X^{J}X^{K}$
correspond to the size of the two- and four-cycles of the Calabi-Yau
threefold respectively. Here $C_{IJK}$ are the intersection numbers of the
threefold. In the Calabi-Yau cases, $n$ is given by the Hodge number $%
h_{(1,1)}$. Some useful relations from very special geometry are 
\begin{eqnarray}
&&X^{I}X_{I}=1,\qquad \partial _{i}X_{I}=-\frac{2}{3}G_{IJ}\partial
_{i}X^{J},  \notag \\
&&X_{I}=\frac{2}{3}G_{IJ}X^{J},\qquad X_{I}\partial _{i}X^{I}=X^{I}\partial
_{i}X_{I}=0.  \label{useful}
\end{eqnarray}

In gauged supergravity theories, the supersymmetry transformations of the
gravitino and gauginos in a bosonic background are given by \cite{bcs1} 
\begin{eqnarray}
\delta \psi _{\mu } &=&\left[ \mathcal{D}_{\mu }+\frac{i}{8}X_{I}(\Gamma
_{\mu }{}^{\nu \rho }-4\delta _{\mu }^{\nu }\Gamma ^{\rho })F_{\nu \rho
}^{I}+\frac{1}{2}g\Gamma _{\mu }X^{I}V_{I}\right] \epsilon ,  \notag \\
\delta \lambda _{i} &=&\left[ -\frac{1}{4}G_{IJ}\Gamma ^{\mu \nu }F_{\mu \nu
}^{J}+\frac{3i}{4}\Gamma ^{\mu }\partial _{\mu }X_{I}+\frac{3i}{2}gV_{I}%
\right] \partial _{i}X^{I}\epsilon .  \label{gst}
\end{eqnarray}
Here $\mathcal{D}_{\mu }$ is the fully gauge and gravitationally covariant
derivative\footnote{%
We use the metric $\eta ^{ab}=(-,+,+,+,+)$, $\{\Gamma ^{a},\Gamma
^{b}\}=2\eta ^{ab}$, $\nabla _{\mu }=\partial _{\mu }+\frac{1}{4}\omega
_{\mu ab}\Gamma ^{ab}$, $\omega _{\mu ab}$ is the spin connection, and $%
\Gamma ^{a_{1}a_{2}\cdots a_{n}}=\Gamma ^{\lbrack {a_{1}}}\Gamma
^{a_{2}}\cdots \Gamma ^{a_{n}]}$, where antisymmetrization is taken with
unit weight.}, 
\begin{equation}
\mathcal{D}_{\mu }\epsilon =\left[ \nabla _{\mu }-\frac{3i}{2}gV_{I}A_{\mu
}^{I}\right] \epsilon .
\end{equation}
The ungauged theory is obtained in the (smooth) limit $g\rightarrow 0$.

\section{General Solutions Admitting Killing Spinors}

\label{bpssol}

The Einstein equations with positive cosmological constant admit de~Sitter
space as vacuum solution, with a metric that can be written in the form 
\begin{equation}
ds^{2}=-(1-g^{2}r^{2})dt^{2}+\frac{dr^{2}}{1-g^{2}r^{2}}+r^{2}d\Omega
_{3}^{2},  \label{dSstatic}
\end{equation}
where $d\Omega _{3}^{2}$ is the standard metric on the unit three-sphere.
Denoting by $d\vec{x}^{2}$ the flat line element in four dimensions, (\ref
{dSstatic}) can be rewritten in the cosmological form 
\begin{equation}
ds^{2}=-dt^{2}+e^{2gt}d\vec{x}^{2},  \label{dsm}
\end{equation}
which is related to the above metric by changing the coordinates according
to 
\begin{equation}
r^{\prime }=re^{gt},\quad dt=dt^{\prime }-\frac{gr^{\prime }}{1-g^{2}{%
r^{\prime }}^{2}}dr^{\prime },  \notag
\end{equation}
and then dropping the primes. Multi-centered black hole solutions asymptotic
to (\ref{dsm}) were found in \cite{london,jw1}. These solutions are given by

\begin{eqnarray}
ds^2 &=& -e^{-4U}dt^{2}+e^{2U}e^{2gt}d\vec{x}^{2}, \label{mkl} \\
A_t^I &=& e^{-2U}X^{I}, \\
X_I &=& \frac 13 e^{-2U}H_I, \\
H_I &=& 3V_{I}+\sum_{i=1}^N\frac{q_{Ii}}{|\vec{x}_i-\vec{x}|^2}e^{-2gt},
\end{eqnarray}
where the $q_{Ii}$ are related to the electric charges. These solutions
resemble those of the ungauged theory \cite{sabra1} but with the spatial
coordinates modified by an exponential time-dependent factor.

Under the change of variables

\begin{equation}
r^{\prime }=re^{gt},\quad dt=dt^{\prime }-\frac{gr^{\prime }e^{6U}}{1-g^{2}{%
r^{\prime }}^{2}e^{6U}}dr^{\prime}, \notag
\end{equation}

we get (after dropping the primes) for the metric

\begin{equation*}
ds^2 = -e^{-4U}(1-g^2r^2e^{6U})dt^2 + \frac{e^{2U}dr^2}{1-g^2r^2e^{6U}} +
e^{2U}r^2d\Omega_3^2,
\end{equation*}

which is the de~Sitter version of the solutions found in \cite{bcs1}.

From the preceding analysis it is clear that it might be advantageous to
analyze more general charged black hole solutions in the time-dependent
cosmological frame. Therefore, our starting point is the ansatz

\begin{equation}
ds^2 = -e^{-4U}(dt+G(t)w_m dx^m)^2+e^{2U}F^2(t)d\vec{x}^2,  \label{magic}
\end{equation}

where $U=U(x,t)$, $w_{m}=w_{m}(x)$, and $G$, $F$ are time-dependent
functions. The F\"{u}nfbein and its inverse for the metric (\ref{magic}) can
be taken as

\begin{align}
e_t^{\ 0} & = e^{-2U},\quad e_m^{\ 0}=Ge^{-2U}w_m,  \notag \\
e_t^{\ a} & = 0,\quad e_m^{\ a}=e^U F\delta_m^a,  \notag \\
e_{\ 0}^t & = e^{2U},\quad e_{\ a}^t=-\frac GF e^{-U}w_m\delta_a^m,  \notag
\\
e_{\ 0}^m & = 0,\quad e_{\ a}^m=\frac1F e^{-U}\delta_a^m.  \label{funf}
\end{align}

For the spin connection one obtains

\begin{align}
\omega_t^{0a} & = \frac{e^{-3U}}{F}\delta^{am}[(-\dot{G}+2G\dot{U})w_m-
2\partial_m U],  \notag \\
\omega_n^{0a} & = e^{3U}(\dot{F}+F\dot{U})\delta_n^a-\frac{G}{2F}e^{-3U}
\delta^{am}(\partial_n w_m-\partial_m w_{n})  \notag \\
& + \frac{e^{-3U}G}{F}\delta^{am}w_n[(-\dot{G}+2G\dot{U})w_m-2\partial_mU],
\\
\omega_t^{ab} & = -\frac{G}{2F^2}e^{-6U}\delta^{nb}\delta^{ma}(\partial_n
w_m-\partial_m w_n),  \notag \\
\omega_n^{ab} & = (\delta^{mb}\delta_n^a-\delta^{ma}\delta_n^b) \left[%
\partial_m U-w_m G\left(\dot{U}+\frac{\dot{F}}{F}\right)\right]  \notag \\
& -\frac{G^2}{2F^2}e^{-6U}w_n\delta^{pb}\delta^{ma}(\partial_p
w_m-\partial_m w_p).
\end{align}

We will look for configurations which admit Killing spinors. The Killing
spinor equations are obtained by setting to zero the supersymmetry
transformations of the Fermi fields (in a bosonic background) for the choice 
$\Gamma _{0}\epsilon =i\epsilon $. From the time component of the gravitino
supersymmetry transformation, one thus obtains

\begin{align}
\delta\psi_t & = \partial_t\epsilon - \frac i2\omega_t^{0a}\Gamma_a \epsilon
- \frac{1}{4F^2}e^{-6U}\delta_a^m\delta_b^n(G(\partial_n w_m - \partial_m
w_n)+e^{2U}X_I F_{mn}^I)\Gamma^{ab}\epsilon  \notag \\
& -\frac{i}{2F}e^{-U}X_I F_{tm}^I\delta_a^m\Gamma^a\epsilon + \frac i2
ge^{-2U}X^I V_I\epsilon - \frac{3i}{2}gV_I A_t^I\epsilon.  \label{tgt}
\end{align}

Using the relation

\begin{equation}
\Gamma^{ab} = -\frac i2\epsilon^{abcd0}\Gamma_{cd}\Gamma_0,
\end{equation}

we then get from the vanishing of (\ref{tgt}) the equations

\begin{align}
(\partial_t+\frac i2 ge^{-2U}X^I V_I-\frac{3i}{2}gV_I A_t^I)\epsilon & = 0, 
\notag \\
-e^{-2U}[(-\dot{G}+2G\dot{U})w_m-2\partial_m U]-X_I F_{tm}^I & = 0,
\label{tg} \\
(X_I F_{mn}^I)^{-} & = (\partial_n Q_m-\partial_m Q_n)^{-},  \notag
\end{align}

where $Q_n \equiv Ge^{-2U}w_n$ and $F_{mn}^{-}=F_{mn}-\,^{\star}\!F_{mn}$.
Note that the second equation in (\ref{tg}) can be rewritten as

\begin{equation*}
X_I F_{tm}^I=\partial_t Q_m-\partial_m e^{-2U}.
\end{equation*}

Thus, using also the relations $X_I\partial X^I=0$, $X_I X^I=1$, we deduce
the following solution for the gauge fields,

\begin{equation*}
A_{m}^{I}=e^{-2U}w_{m}X^{I},\quad A_{t}^{I}=e^{-2U}X^{I},
\end{equation*}

and therefore the first and the third relation in (\ref{tg}) reduce to

\begin{eqnarray}
(\partial _{t}-ige^{-2U}V_{I}X^{I})\epsilon &=&0,  \label{ff1} \\
X_{I}F_{mn}^{I} &=&(\partial _{n}Q_{m}-\partial _{m}Q_{n}).  \label{ff2}
\end{eqnarray}

We turn now to the spatial component of the gravitino supersymmetry
transformation, which for our ansatz reads

\begin{align}
\delta\psi_m & = \partial_m\epsilon + \frac14(\omega_m^{ab}\Gamma_{ab}+2i
\omega_m^{a0}\Gamma_a)\epsilon + \frac i8 X_I(\Gamma_m^{\
np}F_{np}^I+2\Gamma_m^{\ nt}F_{nt}^I)\epsilon  \notag \\
& -\frac i2 X_I(\Gamma^nF_{mn}^I+\Gamma^t F_{mt}^I)\epsilon + \frac12
g\Gamma_m X^I V_I\epsilon - \frac{3i}{2}gV_I A_m^I\epsilon.
\end{align}

By substituting the relations of the gauge fields, we find that the $%
\Gamma_a $ and $\Gamma_{ab}$ terms vanish provided

\begin{align}
(\partial_m w_n-\partial_n w_m)^{-} & = 0,  \label{cond1} \\
-ie^{2U}(\dot{F}+F\dot{U})+gX^I V_IF & = 0,  \label{cond2} \\
GF^2 & =1.  \label{cond3}
\end{align}

Furthermore, we obtain the differential equation

\begin{equation}
\left(\partial_m+\partial_m U\right)\epsilon = 0.
\end{equation}

Using (\ref{cond2}), Eq.~(\ref{ff1}) becomes

\begin{equation*}
(\partial_t+\frac{\dot{F}}{F}+\dot{U})\epsilon = 0.
\end{equation*}

Therefore the configuration found admits the Killing spinor

\begin{equation}
\epsilon = \frac{e^{-U}}{F}\epsilon_0,
\end{equation}

where $\epsilon_0$ is a constant spinor satisfying the constraint $%
\Gamma_0\epsilon_0=i\epsilon_0$.

In order to fix the scalar fields and therefore the gauge fields in terms of
space-time functions, we turn to the gauge field equations of motion. These
are given by

\begin{equation}
\partial_{\nu}(eG_{IJ}g^{\mu\rho}g^{\nu\sigma}F_{\rho\sigma}^J)= \frac{1}{16}%
C_{IJK}\epsilon^{\mu\nu\rho\sigma\kappa}F_{\nu\rho}^J F_{\sigma\kappa}^{K}.
\end{equation}

After some lengthy calculations one finds

\begin{equation}
e^{2U}X_I=\frac13 H_I,
\end{equation}

with the harmonic functions $H_I=h_I+\frac{q_I}{r^2}G$.

Finally one can fix $h_{I}$ and $G$ by turning to the equation obtained from
the vanishing of the supersymmetry variation for the gauginos. This yields

\begin{equation*}
\left(-G_{IJ}\Gamma^{\mu\nu}F_{\mu\nu}^J+3i\Gamma^{\mu}\partial_{\mu}X_I+
6igV_I\right)\partial_i X^I\epsilon = 0.
\end{equation*}

Requiring the term independent of the $\Gamma^a$ to be zero gives

\begin{equation}
(2igV_I+e^{2U}\partial_t X_I)\partial_i X^I\epsilon = 0.  \label{constterm}
\end{equation}

The vanishing of the $\Gamma^{ab}$ and $\Gamma^a$ terms yields

\begin{align*}
\left(G_{IJ}\partial_i X^I F_{mn}^J+\frac{3G}{2}e^{-2U}(\partial_m X_I
w_n-\partial_n X_I w_m)\partial_i X^I\right)^- & =0, \\
\left(-\frac1F G_{IJ}e^U F_{tm}^J-\frac{3G}{2F}e^{-U}w_m\partial_tX_I+ \frac{%
3}{2F}e^{-U}\partial_m X_I\right)\partial_i X^I\epsilon & = 0.
\end{align*}

The above equations are satisfied for our ansatz. Relation (\ref{constterm})
fixes $G(t)=e^{2igt}$ and $h_I = 3V_I$.\newline
If we define the rescaled coordinates

\begin{equation}
Y_I=e^{2U}X_I, \qquad Y^I=e^U X^I,
\end{equation}

then the underlying very special geometry implies that

\begin{equation}
Y_IY^I=e^{3U}X_IX^I=e^{3U}=\mathcal{V}(Y)=\frac13C_{IJK}Y^IY^JY^K.
\end{equation}

To summarize, we have found a black hole solution which breaks half of
supersymmetry, given by

\begin{eqnarray}
ds^2 &=& -\mathcal{V}^{-4/3}(Y)(dt+e^{2igt}w_m dx^m)^2+\mathcal{V}^{2/3}(Y)
e^{-2igt}d\vec{x}^2,  \notag \\
\frac12 C_{IJK}Y^JY^K &=& H_I = h_I+\frac{q_I}{r^2}e^{2igt},  \notag \\
A_t^I &=& \frac{Y^I}{\mathcal{V}(Y)},  \label{finalmetric} \\
A_m^I &=& e^{2igt}w_m A_t^I,  \notag \\
(\partial_m w_n-\partial_n w_m)^{-} &=& 0.  \notag
\end{eqnarray}

Clearly the solution depends very much on the choice of the model under
consideration. Explicit solutions can be found for the case with no vector
multiplets, the $STU$ model and the models considered in \cite{gaida}. The
only additional feature here is the dependence of the fields on time. At
first sight, the complex time-dependence occurring in (\ref{finalmetric})
seems awkward. This is certainly no problem in de~Sitter space, where the
coupling constant $g$ is taken to be imaginary. Then all fields are real,
and one can also have multi-centered black holes\footnote{For the
case without vector multiplets, these have been obtained in \cite
{ks3}. For further discussion see \cite{jw1}.}. However, for the
one-centered de~Sitter solution\footnote{%
Which is obtained from (\ref{finalmetric}) by the analytical continuation
$g\rightarrow ig$.}, as we shall see in the next section, one can go to a set
of new coordinates in which the metric depends only on $g^2$, whereas
the gauge fields and scalars are independent of $g$, and
therefore the anti-de~Sitter rotating solution can then be simply obtained
by replacing $g^{2}$ with $-g^{2}$ \cite{jw1}, so that all fields are real.

\section{Solutions with Rotational Symmetry}

\label{rotsymm}

Using the above general ansatz one can construct models with rotational
symmetry. If we introduce spherical coordinates by

\begin{eqnarray}
x^1+ix^2 &=& r\sin\theta e^{i\phi}, \quad x^3+ix^4 = r\cos\theta e^{i\psi}, 
\notag \\
d\vec{x}^2 &=& dr^2+r^2d\Omega^2 = dr^2 + r^2(d\theta^2+\sin^2\theta d\phi^2
+\cos^2\theta d\psi^2),  \label{sc}
\end{eqnarray}

and specialize to solutions with rotational symmetry in two orthogonal
planes, i.~e.

\begin{equation}
w_{\phi}=w_{\phi}(r,\theta), \qquad w_{\psi}=w_{\psi}(r,\theta), \qquad
w_r=w_{\theta }=0,
\end{equation}

then the self-duality condition of the field strength of $w$ implies \cite
{chamsabra98}

\begin{align}
\partial_r w_{\phi}+\frac{\tan\theta}{r}\partial_{\theta}w_{\psi} & = 0, 
\notag \\
\partial_{\theta}w_{\phi}-r\tan\theta\partial_r w_{\psi} & = 0.
\end{align}

For an asymptotically decaying solution, one finds\footnote{%
This amounts in cartesian coordinates to $w_1 = \frac{\alpha x^2}{r^4},
\quad w_2 = -\frac{\alpha x^1}{r^4}, \quad w_3 = -\frac{\alpha x^4}{r^4},
\quad w_4 = \frac{\alpha x^3}{r^4}$.}

\begin{equation}
w_{\phi}=-\frac{\alpha}{r^2}\sin^2\theta, \qquad w_{\psi}=\frac{\alpha}{r^2}
\cos^2\theta.
\end{equation}

If we analytically continue to de~Sitter space, $g\rightarrow ig$, the
solution (\ref{finalmetric}) reads

\begin{equation}
ds^{2}=-e^{-4U}(dt-e^{-2gt}\frac{\alpha }{r^{2}}\sin ^{2}\theta d\phi
+e^{-2gt}\frac{\alpha }{r^{2}}\cos ^{2}\theta d\psi
)^{2}+e^{2U}e^{2gt}(dr^{2}+r^{2}d\Omega ^{2}).  \label{adssol}
\end{equation}

Introducing the new coordinates $(r^{\prime},t^{\prime},\phi^{\prime},
\psi^{\prime})$ according to

\begin{equation}
r^{\prime} = re^{gt}, \quad dt=dt^{\prime}+f(r^{\prime})dr^{\prime},
\label{coordtransf}
\end{equation}

\begin{equation}
d\phi = d\phi^{\prime}+h(r^{\prime})dr^{\prime}, \quad
d\psi = d\psi^{\prime}-h(r^{\prime})dr^{\prime},
\notag
\end{equation}

where

\begin{eqnarray*}
f(r^{\prime}) &=& \frac{-gr^{\prime}\left(e^{6U}-\frac{\alpha^2} {{r^{\prime}%
}^6}\right)}{1-g^2{r^{\prime}}^2 e^{6U}+\frac{g^2\alpha^2} {{r^{\prime}}^4}},
\\
h(r^{\prime}) &=& \frac{\alpha g}{{r^{\prime}}^3 \left(1-g^2 {%
r^{\prime}}^2 e^{6U} + \frac{g^2\alpha^2}{{r^{\prime}}^4}\right)},
\end{eqnarray*}

we obtain (after dropping the primes) the metric in stationary form,

\begin{eqnarray}
ds^2 &=& e^{2U}g^2r^2dt^2-e^{-4U}(dt-\frac{\alpha}{r^2}\sin^2\theta d\phi + 
\frac{\alpha}{r^2}\cos^2\theta d\psi)^2  \notag \\
&& +e^{2U}\left(\frac{dr^2}{1-g^2r^2 e^{6U}+\frac{g^2\alpha^2}{r^4}}
+r^2d\Omega^2\right).  \label{finalds}
\end{eqnarray}

For the gauge fields in the new coordinates, one gets

\begin{equation}
A^I_{\phi} = -e^{-2U}X^I\frac{\alpha}{r^2}\sin^2\theta, \quad A^I_{\psi} =
e^{-2U}X^I\frac{\alpha}{r^2}\cos^2\theta, \quad A^I_t = e^{-2U}X^I,
\end{equation}

so they are now time-independent. As a special case, one obtains the
solution of the five-dimensional Einstein-Maxwell theory considered in \cite
{ks3}.

By replacing $g^2$ with $-g^2$, one can now obtain the rotating AdS black
holes for all supergravity theories with vector multiplets, which is given by

\begin{eqnarray}
ds^2 &=& -e^{2U}g^2r^2dt^2-e^{-4U}(dt-\frac{\alpha}{r^2}\sin^2\theta d\phi + 
\frac{\alpha}{r^2}\cos^2\theta d\psi)^2  \notag \\
&&+e^{2U}\left(\frac{dr^2}{1+g^2r^2e^{6U}-\frac{g^2\alpha^2}{r^4}}+
r^2d\Omega^2\right).  \label{finalads}
\end{eqnarray}

As an example, consider the so-called $STU=1$ model\footnote{We apologize
for using the same symbol for one of the moduli and the
function appearing in the metric, but the meaning should be clear from the
context.} $(X^0=S, X^1=T, X^2=U)$. Using $e^{2U}X_{I}=H_I/3$,
we get the following equations for the moduli fields,

\begin{eqnarray}
e^{2U}TU &=&H_{0}=h_{0}+\frac{q_{0}}{r^{2}},  \notag \\
e^{2U}SU &=&H_{1}=h_{1}+\frac{q_{1}}{r^{2}},  \label{j} \\
e^{2U}ST &=&H_{2}=h_{2}+\frac{q_{2}}{r^{2}}.  \notag
\end{eqnarray}

Equation (\ref{j}) together with the fact that $STU=1$ implies that the
metric and the moduli fields are given by

\begin{equation}
e^{6U} = H_0H_1H_2,
\end{equation}

and

\begin{equation}
S=\left(\frac{H_1H_2}{H_0^2}\right)^{\frac13}, \qquad T=\left(\frac{H_0H_2}
{H_1^2}\right)^{\frac13}, \qquad U=\left(\frac{H_0H_1}{H_2^2}\right)^{\frac13}.
\end{equation}

For anti-de~Sitter supergravity, one obtains the charged rotating
one-centered solution of the $STU=1$ model,

\begin{eqnarray}
ds^2 &=& -(H_0H_1H_2)^{\frac13}g^2r^2dt^2-(H_0H_1H_2)^{-\frac23}(dt- \frac{%
\alpha}{r^2}\sin^2\theta d\phi + \frac{\alpha}{r^2}\cos^2\theta d\psi)^2 
\notag \\
&& +(H_0H_1H_2)^{\frac13}\left(\frac{dr^2}{1+g^2r^2(H_0H_1H_2)- \frac{%
g^2\alpha^2}{r^4}}+r^2d\Omega^2\right). \label{solSTU=1}
\end{eqnarray}

The gauge fields are then given by

\begin{equation*}
A_{t}^{I}=H_{I}^{-1},\quad A_{m}^{I}=w_{m}A_{t}^{I}.
\end{equation*}

\bigskip

\bigskip For de Sitter space, one can obtain multi-centered solutions where $%
e^{6U}=H_{0}H_{1}H_{2}$ in (\ref{adssol}) with

\begin{eqnarray}
H_{0} &=& 3V_0 + \sum_{i=1}^{N}\frac{q_{0i}}{|\vec{x}_{i}-\vec{x}|^{2}}%
e^{-2gt},  \notag \\
H_{1} &=& 3V_1 + \sum_{i=1}^{N}\frac{q_{1i}}{|\vec{x}_{i}-\vec{x}|^{2}}%
e^{-2gt}, \\
H_{2} &=& 3V_2 + \sum_{i=1}^{N}\frac{q_{2i}}{|\vec{x}_{i}-\vec{x}|^{2}}%
e^{-2gt}.  \notag
\end{eqnarray}

\section{Physical Discussion}

\label{physdisc}

\subsection{Horizons, Temperature, Entropy and Closed Timelike Curves}

To discuss the conditions for the existence of horizons for the
solution (\ref{finalads}), we limit ourselves to the case without vector
multiplets, where one has
\begin{equation}
e^{2U}=1+\frac{q}{r^{2}}.
\end{equation}
From (\ref{finalads}), we see that horizons occur whenever 
\begin{equation}
1+g^{2}r^{2}e^{6U}-\frac{g^{2}\alpha ^{2}}{r^{4}}=0,  \label{condhor}
\end{equation}
which yields a third order equation for $r^{2}$. We define the dimensionless
parameters 
\begin{equation}
a\equiv \alpha g^{3},\quad \rho =qg^{2},
\end{equation}
as well as the critical rotation parameter 
\begin{equation}
a_{c}^{2}=\rho ^{2}+\frac{2}{3}\rho +\frac{2}{27}-\frac{2}{27}(1+6\rho
)^{3/2}.
\end{equation}
For $a^{2}<a_{c}^{2}$ one finds no horizons, whereas for $\rho ^{2}\geq
a^{2}>a_{c}^{2}$ one gets an inner and an outer horizon. For
$a^{2}>\rho ^{2}$ the inner horizon disappears, and we have only one
horizon. For $a^{2}=a_{c}^{2}$, the two horizons coalesce, and one is left
with an extremal black hole with zero Hawking temperature.

As mentioned in the introduction, the solution (\ref{finalads}) admits
closed timelike curves outside the horizon. To see this, we note that
the vector $\partial_{\phi} - \partial_{\psi}$, which has closed orbits,
becomes timelike for
\begin{equation}
r^2e^{6U} - \frac{\alpha^2}{r^4} < 0. \label{VLS}
\end{equation}
One easily sees that this can occur outside the horizon. In
particular, for the case without vector multiplets, (\ref{VLS})
is satisfied for $r^2 < r_L^2 \equiv \alpha^{2/3} - q$, and one has
$r_L > r_+$, where $r_+$ is the location of the horizon.
This causality violation already appears in the ungauged case $(g=0)$,
i.~e.~in the BMPV black hole \cite{BMPV}, where closed timelike curves
exist outside the horizon if $\alpha^2 > q^3$ ("over-rotating" black hole).
Various aspects of this "time machine" have been studied
in \cite{gibbherd,herd}. Among other things, the authors of \cite{gibbherd}
showed that causal geodesics cannot penetrate the $r=r_+$ surface
for the over-rotating solution (repulson-like behaviour), and so
the exterior region is geodesically
complete (with respect to causal geodesics).
Furthermore, in \cite{herd}, it was shown that upon lifting
of the over-rotating BMPV black hole to a solution of
type IIB string theory and passing
to the universal covering space, the causal anomalies disappear.
A detailed discussion of the issues related with the causality
violation encountered for the solution (\ref{finalads}) will
be presented in \cite{cks}. We mention here briefly some of the results:
Also in the gauged case, the exterior region is geodesically complete;
no causal geodesics cross the $r=r_+$ surface. The five-dimensional
solution for the $STU=1$ model can be lifted to a solution of
type IIB supergravity in ten dimensions using the Kaluza-Klein
ansatz of \cite{tenauthors}. One then obtains a rotating D3-brane
wrapping $S^3$. The resulting ten-dimensional configuration still has closed
timelike curves, so in our case no resolution of causal anomalies
in higher dimensions appears.

Note that the naively computed temperature and Bekenstein-Hawking
entropy are given by
\begin{eqnarray}
T &=& \frac{r_+^3}{4\pi\sqrt{e^{6U(r_+)}r_+^6 - \alpha^2}}\frac{d}{dr}
\left[1 + g^2r^2e^{6U} - \frac{g^2\alpha^2}{r^4}\right]_{|r=r_+}, \nonumber \\
S_{BH} &=& \frac{A_{hor}}{4G} = \frac{\pi^2}{2G}\sqrt{e^{6U(r_+)}r_+^6
- \alpha ^2}.
\end{eqnarray}
These quantities are imaginary and hence become senseless. For the
over-rotating ungauged case, it was argued in \cite{gibbherd} that the
effective entropy should be considered to be zero. The reason for this
was the vanishing of the absorption cross section for a massless
Klein-Gordon field, which is a measure of the horizon area.
Also for the solution (\ref{solSTU=1}) of the $STU=1$ model,
the absorption probability for a Klein-Gordon field
vanishes \cite{cks}, indicating that the effective temperature and entropy
are both zero.

The causality violation in the spacetime (\ref{finalads}) should
translate on the dual CFT side into loss of unitarity
(relation between macroscopic causality in the AdS bulk and
microscopic unitarity in the boundary CFT). For the rotating BMPV
black hole, which corresponds to a D1-D5-Brinkmann wave system
in type IIB string theory \cite{herd}, one has a description in terms
of an ${\cal N}=4$ two-dimensional superconformal field theory.
The causality bound $\alpha^2 = q^3$ then corresponds to the
unitarity bound in this CFT \cite{herd}.
Probably the appearance of closed timelike curves can be avoided
in the nonsupersymmetric generalization of (\ref{finalads}),
because there would be one more parameter (the nonextremality parameter
$\mu$) available. It should then be possible to shift the region of closed
timelike curves beyond the horizon by choosing $\mu$ sufficiently large.
It would be interesting to determine the resulting unitarity bound
for $D=4$, ${\cal N}=4$ SYM theory, which would be obtained by requiring
the absence of closed timelike curves outside the horizon.
This unitarity bound could then be compared to the one coming from
the superconformal algebra of the corresponding CFT. The classification
of unitary representations of superconformal algebras typically
implies inequalities on the conformal weights and
$R$-charges \cite{mack}. On passing from four-dimensional Minkowski
space to $\bR\times S^3$ (where our CFT lives, cf.~below), eigenvalues
of the generator of dilatations $D$ translate into eigenvalues of
the Hamiltonian $H$. Thus the inequalities mentioned above
yield lower bounds on the black hole mass $M$, and these lower bounds are
given in terms of the $R$-charges. As $M$ typically is proportional to
the nonextremality parameter $\mu$, it is clear that by turning on a
sufficiently large $\mu$, the unitarity bound can be satisfied.
Clearly the nonextremal generalization of (\ref{finalads})
is not yet known, but there should exist a general solution which contains
both (\ref{finalads}) and the nonextremal uncharged Kerr-AdS black holes
found in \cite{hht}.

\subsection{Stress Tensor of the Dual CFT}

In computing quantities like the Euclidean action, which yields the
thermodynamical potential relevant to the various ensembles, one usually
encounters infrared divergences, which are regularized by subtracting a
suitably chosen background. Such a procedure, however, in general is not
unique; in some cases the choice of reference background is ambiguous,
e.~g.~for hyperbolic AdS black holes \cite{vanzo97,ejm99}. Recently, in
order to regularize such divergences, a different procedure has been
proposed \cite{skenderis1,balasubramanian99,ejm99}\footnote{Cf.~also
\cite{skenderis2}.}. This technique
was inspired by the
AdS/CFT correspondence, and consists in adding suitable counterterms $I_{ct}$
to the action. These counterterms are built up with curvature invariants of
a boundary $\partial \mathcal{M}$ (which is sent to infinity after the
integration), and thus obviously they do not alter the bulk equations of
motion. This kind of procedure, which will also be used in the present
paper, has the advantage of being free of ambiguities, which, on the
contrary, are present in the traditional approach in some particular cases,
like mentioned above.

In the following discussion, we only consider the case without
vector multiplets, and leave the general case for a future publication.
The Euclidean action for Einstein-Maxwell-AdS gravity has the
form\footnote{This action is related to that of \cite{london} by rescaling
the gauge fields of \cite{london} by $\frac{1}{2\sqrt 3}$.}
\begin{eqnarray}
I &=& I_{bulk}+I_{surf}+I_{ct}  \notag \\
&=& -\frac1{16\pi G}\int_{\mathcal{M}}\!d^5x \sqrt{g}\left[R+12g^2-\frac{1}{12}
F_{\mu\nu}F^{\mu\nu} + \frac1{108\sqrt g}\epsilon^{\mu\nu\rho\sigma\lambda}
F_{\mu\nu}F_{\rho\sigma}A_{\lambda}\right] \nonumber \\
& & -\frac1{8\pi G}\int_{\partial\mathcal{M}}\!d^4x\ \sqrt{h}\ K+I_{ct}\:.
\label{euclaction}
\end{eqnarray}
The second term is the Gibbons-Hawking boundary term, where $h_{ab}$ is the
boundary metric, and $K$ denotes the trace of the extrinsic curvature $K_{ab}$.

The action (\ref{euclaction}) differs from the familiar one by the presence
of the last term, which contains all surface counterterms needed to insure
the convergence of the integrals. In five dimensions it reads \cite{ejm99} 
\begin{equation}
I_{ct}=\frac{1}{8\pi G}\int_{\partial \mathcal{M}}\!d^4 x\sqrt h\left[3g+
\frac{\mathcal{R}}{4g}\right], \label{actionCT}
\end{equation}
$\mathcal{R}$ denoting the Ricci scalar of the boundary.
Going back to
Lorentzian signature, one can now construct a divergence free stress tensor
given by 
\begin{equation}
T^{ab}=\frac{2}{\sqrt{-h}}\frac{\delta I}{\delta h_{ab}}=\frac{1}{8\pi G}
\left[ K^{ab}-h^{ab}K-3gh^{ab}+\frac{G^{ab}}{2g}\right],
\label{stresstensgen}
\end{equation}
where $G^{ab}$ denotes the Einstein tensor built up from $h_{ab}$. If we
choose $\partial \mathcal{M}$ to be a four-surface of fixed $r$, then the
metric on the manifold upon which the dual CFT resides is defined by 
\begin{equation}
\gamma _{ab}=\lim_{r\rightarrow \infty }\frac{1}{g^{2}r^{2}}h_{ab}.
\end{equation}
In our case, this yields 
\begin{equation}
\gamma_{ab}dx^adx^b=-dt^2+\frac{1}{g^2}d\Omega_3^2, \label{boundmetr}
\end{equation}
where $d\Omega_3^2$ is the standard metric on the unit $S^3$. We see that
the dual CFT is defined on the (nonrotating) manifold $\bR\times S^{3}$. The
field theory's stress tensor $\hat{T}^{ab}$ is related to the one in (\ref
{stresstensgen}) by the rescaling \cite{myers}
\begin{equation}
\sqrt{-\gamma }\gamma _{ab}\hat{T}^{bc}=\lim_{r\rightarrow \infty }\sqrt{-h}%
h_{ab}T^{bc},
\end{equation}
which amounts to multiplying all expressions for $T_{ab}$ given below by
$g^2r^2$ before taking the limit $r\rightarrow \infty$.

Before we proceed, a short comment is in order. In five dimensions,
in general one encounters also logarithmic divergences
in the computation of the Euclidean action.
These divergences, which cannot be removed by adding local counterterms,
are related to the Weyl anomaly of the dual CFT \cite{skenderis1}.
In our case, the divergent term reads \cite{skenderis1}
\begin{equation}
I_{log} = -\frac{\ln \epsilon}{64\pi Gg^3}\int_{{\cal M}}d^4 x\sqrt{\gamma}
          \left[{\cal R}_{ab}{\cal R}^{ab} - \frac13 {\cal R}^2\right],
\end{equation}
where $\epsilon$ denotes a bulk infrared cutoff. For the metric
(\ref{boundmetr}), one finds that this term vanishes, so there is no
conformal anomaly in the dual field theory. In principle, an additional
logarithmically divergent term can appear in the Euclidean action
(\ref{euclaction}) due to the Maxwell gauge field $A_{\mu}$ \cite{marika}.
However, for the black holes considered here, this term also
vanishes\footnote{In our case, $F_{\mu\nu}F^{\mu\nu}$ falls off as
$r^{-6}$ for $r\to\infty$, whereas a logarithmic divergence
would come from a term of order $r^{-4}$ \cite{marika}.}.

We now come to the computation of the stress tensor (\ref{stresstensgen}).
Using (\ref{finalads}), one obtains
\begin{eqnarray}
8\pi GT_{tt} &=& \frac{3}{8gr^2}(8\rho+1) + \mathcal{O}(r^{-4}),  \notag \\
8\pi GT_{t\phi} &=& \frac{2g\alpha}{r^2}\sin^2\theta + \mathcal{O}(r^{-4}), 
\notag \\
8\pi GT_{t\psi} &=& -\frac{2g\alpha}{r^2}\cos^2\theta + \mathcal{O}(r^{-4}),
\notag \\
8\pi GT_{\phi\phi} &=& \frac{\sin^2\theta}{8g^3r^2}(8\rho+1) +\mathcal{O}%
(r^{-4}),  \label{Tab} \\
8\pi GT_{\psi\psi} &=& \frac{\cos^2\theta}{8g^3r^2}(8\rho+1) +\mathcal{O}%
(r^{-4}),  \notag \\
8\pi GT_{\phi\psi} &=& \mathcal{O}(r^{-4}),  \notag \\
8\pi GT_{\theta\theta} &=& \frac{1}{8g^3r^2}(8\rho+1) + \mathcal{O}(r^{-4}),
\notag
\end{eqnarray}
all other components vanishing. Introducing the unit time-like four vector $%
v = (1,0,0,0)$ as well as the null vectors $w_{\pm} = (1,0,\pm
g/\sin\theta,0)$, $z_{\pm} = (1,0,0,\pm g/\cos\theta)$, one can write the
field theory's stress tensor in the form\footnote{%
A similar form has been obtained for five-dimensional uncharged Kerr-AdS
black holes in \cite{awad}.} 
\begin{eqnarray}
8\pi G \hat{T}^{ab} &=& \frac g8 (8\rho+1)[4v^av^b + \gamma^{ab}]  \notag \\
& & -g^4\alpha\sin\theta[w^a_+w^b_+ - w^a_-w^b_-] +g^4\alpha\cos\theta[%
z^a_+z^b_+ - z^a_-z^b_-].
\end{eqnarray}
One easily checks that this tensor is conserved and traceless.\newline
The stress tensor (\ref{Tab}) can also be used to compute conserved
quantities like mass and angular momenta of the black hole. To do this, we
indicate by $u^{\mu}$ the unit normal vector of a spacelike hyper-surface $^4%
\mathcal{S}_t$ at constant $t$, and by $\Sigma$ the spacelike intersection $%
^4\mathcal{S}_t\cap \partial \mathcal{M}$ embedded in $\partial \mathcal{M}$
with induced metric $\sigma_{ab}$. Then, for any Killing vector field $%
\xi^{\mu}$ there is an associated conserved charge 
\begin{equation}
Q_{\xi}=\int_{\Sigma}\:d^3x\:\sqrt{\sigma}\:u^{\mu}T_{\mu\nu}\xi^{\nu}\:.
\label{charges}
\end{equation}
In this way one gets for the black hole mass 
\begin{equation}
M=Q_{\partial_t} = \frac{3\pi}{32Gg^2}(8\rho+1),
\end{equation}
and for the angular momenta 
\begin{eqnarray}
J_{\phi} &=& Q_{\partial_\phi} = \frac{\alpha\pi}{4G},  \notag \\
J_{\psi} &=& Q_{\partial_\psi} = -\frac{\alpha\pi}{4G}.
\end{eqnarray}
The angular momenta coincide with those of the ungauged case calculated in 
\cite{chamsabra98}.

\section{De~Sitter Case}

\label{desitter}

We turn now to the discussion of the de~Sitter black holes (\ref{finalds}).
These solutions could
be relevant to the proposed duality \cite{hull} between the large $N$ limit of
Euclidean four-dimensional $U(N)$ super Yang-Mills theory and type IIB$^*$
string theory in de~Sitter space.\\
For pure Einstein-Maxwell gravity with cosmological constant, some of the
physical properties of the de~Sitter black holes (\ref{finalds})
have been discussed in \cite{ks3}. The new feature which
arises is the presence of a cosmological horizon at $r=r_c>r_+$, where
$r=r_+$ is the location of the black hole event horizon\footnote{For
the de~Sitter solutions, one can show that at least in the case without
vector multiplets or for the $STU=1$ model, no closed timelike curves
appear in the region between the event horizon and the cosmological horizon.}.
In this section we
will generalize some of the results of \cite{ks3} to the case of additional
vector and scalar fields.

\subsection{Smarr Formula}

To obtain a Smarr-type formula for the de Sitter solutions, from which a
first law of black hole mechanics can be deduced, we proceed along the lines
of \cite{gibbhawk}, where the four-dimensional Kerr-Newman-de~Sitter black
hole was considered. We start from the Killing identity

\begin{equation}
\nabla_{\mu}\nabla_{\nu}K^{\mu} = R_{\nu\rho}K^{\rho} = (F^2_{\nu\rho} -
\frac 16 g_{\nu\rho}F^2)K^{\rho} + G_{IJ}\partial_{\nu}X^I\partial_{\rho}X^J
K^{\rho} + \frac 23 g^2V(X)K_{\nu},  \label{killingid}
\end{equation}

where $K^{\mu }$ is a Killing vector, $\nabla _{\left( \nu \right.
}K_{\left. \mu \right) }=0$. Note that in the second step we used the
Einstein equation of motion following from the action (\ref{action}).
Integrating (\ref{killingid}) on a spacelike hypersurface $\Sigma _{t}$ from
the black hole horizon $r_{+}$ to the cosmological horizon $r_{c}$, using
Gauss' law gives

\begin{equation}
\frac 12\int_{\partial\Sigma_t}\nabla_{\mu}K_{\nu}d\Sigma^{\mu\nu} = \frac 23
g^2\int_{\Sigma_t}V(X)K_{\nu}d\Sigma^{\nu} + \int_{\Sigma_t} (F^2_{\nu\rho}
- \frac 16 g_{\nu\rho}F^2 + G_{IJ}\partial_{\nu}X^I
\partial_{\rho}X^J)K^{\rho}d\Sigma^{\nu}, \label{B1}
\end{equation}

where the boundary $\partial\Sigma_t$ consists of the intersection of $%
\Sigma_t$ with the black-hole and the cosmological horizon,

\begin{equation}
\partial\Sigma_t = S^3(r_+) \cup S^3(r_c).
\end{equation}

In a first step, we apply (\ref{B1}) to the Killing vectors $\partial_{\phi}$
and $\partial_{\psi}$, which we denote by $\tilde{K}^i$, $i = \phi,\psi$.
Using $\tilde{K}^i_{\nu}d\Sigma^{\nu} = 0$, we get

\begin{equation}
\frac{1}{16\pi G}\int_{S^3(r_+)}\nabla_{\mu}\tilde{K}^i_{\nu}
d\Sigma^{\mu\nu} + \frac{1}{16\pi G}\int_{S^3(r_c)}\nabla_{\mu}
\tilde{K}^i_{\nu}d\Sigma^{\mu\nu} =
\frac 1G\int_{\Sigma_t}T_{\nu\rho}\tilde{K}^{i\rho}d\Sigma^{\nu},  \label{C}
\end{equation}

where

\begin{equation}
T_{\nu\rho} = \frac{1}{8\pi}\left(F^2_{\nu\rho} - \frac 14 g_{\nu\rho}F^2 +
G_{IJ}\partial_{\nu}X^I\partial_{\rho}X^J - \frac 12 g_{\nu\rho}
G_{IJ}\partial_{\lambda}X^I\partial^{\lambda}X^J\right)
\end{equation}

is the stress-energy tensor of the vector fields and scalars. The right-hand
side of (\ref{C}) can be interpreted as the angular momentum of the matter
between the two horizons. Thus the second term on the left-hand side of (\ref
{C}) can be regarded as being the total angular momentum, $J_{c}^{i}$,
contained in the cosmological horizon, and the first term on the left-hand
side as the negative of the angular momentum $J_{BH}^{i}$ of the black hole
(cf.~also discussion in \cite{gibbhawk}).\newline
In a second step, we apply (\ref{B1}) to the Killing vector $K=\partial _{t}$.
This yields

\begin{eqnarray}
\lefteqn{\frac{3}{32\pi G}\int_{S^3(r_+)}\nabla_{\mu}K_{\nu}d\Sigma^{\mu\nu} + 
\frac{3}{32\pi G}\int_{S^3(r_c)}\nabla_{\mu}K_{\nu}d\Sigma^{\mu\nu} =} \notag
\\
& & \frac{3g^2}{24\pi G}\int_{\Sigma_t}V(X)K_{\nu}d\Sigma^{\nu} +
\frac{3}{2G}\int_{\Sigma_t}\left(T_{\nu\rho} - \frac 13
Tg_{\nu\rho}\right)K^{\rho}d\Sigma^{\nu}.  \label{D}
\end{eqnarray}

The right-hand side of Eq.~(\ref{D}) represents, respectively, the
contribution of the scalar potential energy and the matter kinetic energy to
the mass within the cosmological horizon. Therefore, \ the second term on
the left-hand side is identified as the total mass $M_{c}$ within the
cosmological horizon, and the first term as the negative of the black hole
mass $M_{BH}$. As in Ref.~\cite{4laws}, the latter can be rewritten by
expressing $K=\partial _{t}$ in terms of the null generator

\begin{equation}
l=\partial _{t}+\Omega _{H}^{\phi }\partial _{\phi }+\Omega _{H}^{\psi
}\partial _{\psi }=\partial _{t}+\Omega _{H}^{i}\tilde{K}^{i}
\end{equation}

of the black hole event horizon. Here, the $\Omega^i_H$ denote the angular
velocities of the horizon. Using the definition of the surface gravity
$\kappa_H$, which, by the zeroth law, is constant on the horizon, one obtains

\begin{equation}
M_{BH} = \frac{3}{16\pi G}\kappa_H A_H + \frac 32\Omega^i_HJ^i_{BH}.
\end{equation}

One therefore gets the Smarr-type formula

\begin{eqnarray}
M_{c} &=& \frac{3}{16\pi G}\kappa_H A_H + \frac 32\Omega_H^{\phi}J_{BH}^{\phi}+
\frac 32\Omega_H^{\psi}J_{BH}^{\psi}+\frac{3g^2}{24\pi G}\int_{\Sigma_t}V(X)
K_{\nu}d\Sigma^{\nu}  \notag \\
&& + \frac{3}{2G}\int_{\Sigma_t}\left(T_{\nu\rho}-\frac13Tg_{\nu\rho}\right)
K^{\rho}d\Sigma^{\nu}.
\end{eqnarray}

\section{Final Remarks}

In this paper we derived a class of general charged black hole solutions to
gauged $N=2$, $D=5$ supergravity coupled to an arbitrary number of abelian
vector multiplets. These black holes preserve half of the supersymmetries.
For imaginary coupling constant of the gravitini (de~Sitter case), we
obtained also multicentered charged rotating solutions.
The AdS black holes
have some interesting properties, e.~g.~we saw that
there is a minimum amount of rotation required to avoid naked singularties,
similar to the four-dimensional case \cite{caldo}. However, closed timelike
curves appear outside the horizon. This causality violation indicates
loss of unitarity in the dual superconformal field theory. It would be
interesting to know the nonsupersymmetric generalizations of the solutions
found here, because a sufficiently large nonextremality parameter would
presumably shift the region of closed timelike curves beyond the horizon.
A detailed discussion of the issues related with the encountered
causality violation will be presented in \cite{cks}.

\section*{Acknowledgements}

D.~K.~was partially supported by MURST and
by the European Commission RTN program
HPRN-CT-2000-00131, in which he is associated to the University of Torino.\\
The authors would like to thank K.~Skenderis for useful comments, and
K.~Behrndt and M.~M.~Caldarelli for discussions.


\begin{thebibliography}{99}
\bibitem{adscft} J.~M.~Maldacena, \emph{The large $N$ limit of
superconformal field theories and supergravity}, Adv.~Theor.~Math.~Phys.~%
\textbf{2} (1998) 231; Int.~J.~Theor.~Phys.~\textbf{38} (1999) 1113;\newline
E.~Witten, \emph{Anti-de~Sitter space and holography},
Adv.~Theor.~Math.~Phys.~\textbf{2} (1998) 253;\newline
S.~S.~Gubser, I.~R.~Klebanov and A.~M.~Polyakov, \emph{Gauge theory
correlators from non-critical string theory}, Phys.~Lett.~\textbf{B428}
(1998) 105;\newline
O.~Aharony, S.~S.~Gubser, J.~M.~Maldacena, H.~Ooguri and Y.~Oz,
\emph{Large $N$ field theories, string theory and gravity},
Phys.~Rept.~\textbf{323} (2000) 183.

\bibitem{hawkpage}  S.~W.~Hawking and D.~N.~Page, \emph{Thermodynamics of
black holes in anti-de~Sitter space}, Commun.~Math.~Phys.~\textbf{87} (1983)
577.

\bibitem{witten}  E.~Witten, \emph{Anti-de~Sitter space, thermal phase
transition, and confinement in gauge theories}, Adv.~Theor.~Math.~Phys.~%
\textbf{2} (1998) 505.

\bibitem{chamblin}  A.~Chamblin, R.~Emparan, C.~V.~Johnson and R.~C.~Myers, 
\emph{Charged AdS black holes and catastrophic holography}, Phys.~Rev.~%
\textbf{D60} (1999) 064018; \emph{Holography, thermodynamics and
fluctuations of charged AdS black holes}, Phys.~Rev.~\textbf{D60} (1999)
104026.

\bibitem{emparan99}  R.~Emparan, \emph{AdS/CFT duals of topological black
holes and the entropy of zero energy states}, JHEP \textbf{9906} (1999) 036.

\bibitem{caldklemm}  M.~M.~Caldarelli and D.~Klemm, \emph{M-theory and
stringy corrections to anti-de~Sitter black holes and conformal field
theories}, Nucl.~Phys.~\textbf{B555} (1999) 157.

\bibitem{awad}  A.~M.~Awad and C.~V.~Johnson, \emph{Holographic stress
tensors for Kerr-AdS black holes}, Phys.~Rev.~\textbf{D61} (2000) 084025; 
\emph{Higher dimensional Kerr-AdS black holes and the AdS/CFT correspondence},
hep-th/0008211.

\bibitem{awad2}  A.~M.~Awad and C.~V.~Johnson, \emph{Scale vs.~conformal
invariance in the AdS/CFT correspondence}, hep-th/0006037.

\bibitem{hht} S.~W.~Hawking, C.~J.~Hunter and M.~M.~Taylor-Robinson,
\emph{Rotation and the AdS/CFT correspondence},
Phys.~Rev.~\textbf{D59} (1999) 064005.

\bibitem{mirjam} M.~Cveti\v{c} and S.~S.~Gubser, \emph{Phases of R charged
black holes, spinning branes and strongly coupled gauge theories}, JHEP 
\textbf{9904} (1999) 024; \emph{Thermodynamic stability and phases of
general spinning branes}, JHEP \textbf{9907} (1999) 010;\newline
M.~M.~Caldarelli, G.~Cognola and D.~Klemm, \emph{Thermodynamics of
Kerr-Newman-AdS black holes and conformal field theories},
Class.~Quant.~Grav.~\textbf{17} (2000) 399;\newline
S.~W.~Hawking and H.~S.~Reall, \emph{Charged and rotating black holes and
their CFT duals}, Phys.~Rev.~\textbf{D61} (2000) 024014;\newline
D.~S.~Berman and M.~K.~Parikh, \emph{Holography and rotating AdS black holes}%
, Phys.~Lett.~\textbf{B463} (1999) 168;\newline
K.~Landsteiner and E.~Lopez, \emph{The thermodynamic potentials of Kerr-AdS
black holes and their CFT duals}, JHEP \textbf{9912} (1999) 020; \\
S.~S.~Gubser and I.~Mitra, \emph{Instability of charged black holes in
anti-de~Sitter space}, hep-th/0009126.

\bibitem{susskind}  J.~Polchinski, L.~Susskind and N.~Toumbas, \emph{%
Negative energy, superluminosity and holography}, Phys.~Rev.~\textbf{D60}
(1999) 084006.

\bibitem{chamsabra99} A.~H.~Chamseddine and W.~A.~Sabra, \emph{Magnetic
strings in five-dimensional gauged supergravity theories},
Phys.~Lett.~\textbf{B477} (2000) 329;

\bibitem{klemmsabra00} D.~Klemm and W.~A.~Sabra, \emph{Supersymmetry of
black strings in $D=5$ gauged supergravities}, Phys.~Rev.~\textbf{D62}
(2000) 024003.

\bibitem{nunez}  J.~M.~Maldacena and C.~Nu\~{n}ez, \emph{Supergravity
description of field theories on curved manifolds and a no go theorem},
hep-th/0007018.

\bibitem{bcs1}  K.~Behrndt, A.~H.~Chamseddine and W.~A.~Sabra, \emph{BPS
black holes in $N=2$ five-dimensional AdS supergravity}, Phys.~Lett.~\textbf{%
B442} (1998) 97;\newline
K.~Behrndt, M.~Cveti\v{c} and W.~A.~Sabra, \emph{Non-extreme black holes of
five-dimensional $N=2$ AdS supergravity}, Nucl.~Phys.~\textbf{B553} (1999)
317.

\bibitem{london}  L.~A.~J.~London, \emph{Arbitrary dimensional cosmological
multi-black holes}, Nucl.~Phys.~\textbf{B434} (1995) 709.

\bibitem{jw1}  J.~T.~Liu and W.~A.~Sabra, \emph{Multi-centered black holes
in gauged $D=5$ supergravity}, hep-th/0010025.

\bibitem{sabra1}  W.~A.~Sabra, \emph{General BPS black holes in five
dimensions}, Mod.~Phys.~Lett.~\textbf{A13} (1998) 239.

\bibitem{gaida}  I.~Gaida, S.~Mahapatra, T.~Mohaupt and W.~A.~Sabra, \emph{%
Black holes and flop transitions in M-Theory on Calabi-Yau threefolds},
Class.~Quant.~Grav.~\textbf{16} (1999) 419.

\bibitem{ks3}  D.~Klemm and W.~A.~Sabra, \emph{Charged rotating black holes
in 5d Einstein-Maxwell-(A)dS gravity}, hep-th/0010200.

\bibitem{chamsabra98} A.~H.~Chamseddine and W.~A.~Sabra, \emph{Metrics
admitting Killing spinors in five dimensions}, Phys.~Lett.~\textbf{B426}
(1998) 36.

\bibitem{BMPV} J.~C.~Breckenridge, R.~C.~Myers, A.~W.~Peet and C.~Vafa,
\emph{D-branes and spinning black holes}, Phys.~Lett.~\textbf{B391}
(1997) 93.

\bibitem{gibbherd} G.~W.~Gibbons and C.~A.~R.~Herdeiro,
\emph{Supersymmetric rotating black holes and causality violation},
Class.~Quant.~Grav.~\textbf{16} (1999) 3619.

\bibitem{herd} C.~A.~R.~Herdeiro,
\emph{Special properties of five dimensional BPS rotating black holes}, 
Nucl.~Phys.~\textbf{B582} (2000) 363. 

\bibitem{cks} M.~M.~Caldarelli, D.~Klemm and W.~A.~Sabra,
\emph{Causality violation and naked time machines in $AdS_5$},
to appear.

\bibitem{tenauthors} M.~Cveti\v{c}, M.~J.~Duff, P.~Hoxha, James T.~Liu, H.~Lu,
J.~X.~Lu, R.~Martinez-Acosta, C.~N.~Pope, H.~Sati, T.~A.~Tran,
\emph{Embedding AdS black holes in ten and eleven dimensions}
Nucl.~Phys.~{\bf B558} (1999) 96.

\bibitem{mack} G.~Mack,
\emph{All unitary ray representations of the conformal group $SU(2,2)$
with positive energy}, Commun.~Math.~Phys.~{\bf 55} (1977) 1;\\
V.~Kac, \emph{Representations of classical Lie superalgebras},
in Lecture Notes in Mathematics, Vol.~676, p.~598 (Springer-Verlag,
Berlin, 1978);\\
V.~Dobrev and V.~Petkova, \emph{All positive energy unitary irreducible
representations of extended conformal supersymmetry},
Phys.~Lett.~{\bf B162} (1985) 127.

\bibitem{vanzo97} L.~Vanzo, \emph{Black holes with unusual topology},
Phys.~Rev.~\textbf{D56} (1997) 6475.

\bibitem{ejm99}  R.~Emparan, C.~V.~Johnson and R.~C.~Myers, \emph{Surface
terms as counterterms in the AdS/CFT correspondence}, Phys.~Rev.~\textbf{D60}
(1999) 104001.

\bibitem{skenderis1} M.~Henningson and K.~Skenderis,
\emph{The holographic Weyl anomaly}, JHEP {\bf 9807} (1998) 023.

\bibitem{balasubramanian99}  V.~Balasubramanian and P.~Kraus, \emph{A stress
tensor for anti-de~Sitter gravity}, Commun.~Math.~Phys.~\textbf{208} (1999)
413.

\bibitem{skenderis2} S.~de Haro, K.~Skenderis and S.~N.~Solodukhin,
\emph{Holographic reconstruction of spacetime and renormalization
in the AdS/CFT correspondence}, hep-th/0002230; \\
K.~Skenderis,
\emph{Asymptotically anti-de Sitter spacetimes and their stress energy
tensor}, hep-th/0010138.

\bibitem{myers}  R.~C.~Myers, \emph{Stress tensors and Casimir energies in
the AdS/CFT correspondence}, hep-th/9903203.

\bibitem{marika} M.~Taylor-Robinson,
\emph{More on counterterms in the gravitational action and anomalies},
hep-th/0002125.

\bibitem{hull} C.~M.~Hull,
\emph{Timelike T duality, de~Sitter space, large $N$ gauge theories and
topological field theory},
JHEP {\bf 9807} (1998) 021.

\bibitem{gibbhawk}  G.~W.~Gibbons and S.~W.~Hawking, \emph{Cosmological
event horizons, thermodynamics, and particle creation}, Phys.~Rev.~\textbf{D15}
(1977) 2738.

\bibitem{4laws}  J.~M.~Bardeen, B.~Carter and S.~W.~Hawking, \emph{The four
laws of black hole mechanics}, Commun.~Math.~Phys.~\textbf{31} (1973) 161.

\bibitem{caldo}  M.~M.~Caldarelli and D.~Klemm, \emph{Supersymmetry of
anti-de~Sitter black holes}, Nucl.~Phys.~\textbf{B545} (1999) 434;\newline
V.~A.~Kosteleck\'{y} and M.~J.~Perry, \emph{Solitonic black holes in gauged
$N=2$ supergravity}, Phys.~Lett.~\textbf{B371} (1996) 191.

\end{thebibliography}
\end{document}